
\magnification=1200
\baselineskip=20pt\vskip 2cm
\centerline{Ground State and Excited States of a Confined Condensed Bose Gas}
\vskip 1cm
\centerline{Alexander L. Fetter}\smallskip
\centerline{\it Departments of Applied Physics and Physics,}
\medskip\centerline{\it  Stanford University, Stanford, CA
94305-4060}
\bigskip
\centerline {Abstract}
{\narrower\smallskip The Bogoliubov approximation is used to study the
ground state and low-lying
excited states of a dilute gas of $N$  atomic bosons held in
an isotropic harmonic potential characterized by  frequency
$\omega$ and oscillator length
$d_0$.  By assumption,   the self-consistent condensate has a macroscopic
occupation number $N_0 >> 1$, with
$N-N_0 << N_0$.  For negative scattering length
$ -|a|$, a simple variational trial function  yields an  estimate for the
critical condensate number
$N_{0\,c} = \big({8\pi/25\sqrt{5}}\,\big)^{1/2}\,(d_0/|a|) \approx
0.671\,(d_0/|a|)$ at the onset of
collapse.  For positive scattering length and large
$N_0 >>d_0/a$, the spherical condensate has a well-defined radius $R >>
d_0$, and the  low-lying excited
states are compressional  waves localized near the surface.
 The frequencies of  the lowest radial modes ($n = 0$) for successive
values of orbital angular
momentum
$l$   form a rotational band  $\omega_{0l} \approx \omega_{00} + {1\over 2}
l(l+1)\,\omega\,(d_0/R)^2$, with $\omega_{00} $ somewhat larger than $\omega$.
\smallskip}
\bigskip
{\narrower\smallskip\noindent PACS numbers: 03.75.Fi, 05.30.Jp, 32.80.Pj,
67.90.+z\smallskip}
\vfill\eject
Recent experimental demonstrations of Bose-Einstein condensation in dilute
confined $^{87}$Rb
[1] and $^7$Li [2] have stimulated  theoretical research into their
physical  properties, based largely
 on the Bogoliubov approximation [3], originally introduced as a model
for bulk  superfluid $^4$He. Although  this simple description of liquid He
has long been familiar, much of
its application to Bose condensed dilute atoms has  involved extensive
numerical analysis [4-6].   In
contrast, Baym and Pethick [7]  provide a more physical  description  of
the confined ground state,
emphasizing the relevant dimensionless parameters for $^{87}$Rb, where the
$s$-wave scattering length $a$ is
positive.  The present work extends  this  picture to include  negative
values of $a$ (as found, for example,
in
$^7$Li), along with the  low-lying excited states for positive $a$.

The Bogoliubov model is most simply understood by considering the familiar
second-quantized field
operators that obey boson commutation relations $[\psi ({\bf
r}),\psi^{\dagger}({\bf r'})] = \delta({\bf r -
r'})$. The dynamics follows from the ``grand-canonical hamiltonian'' operator
$$K \equiv H - \mu N = \int dV \,\psi^{\dagger}(T + U -\mu)\psi + 2\pi a
\hbar^2m^{-1}\int
dV\,\psi^{\dagger}\psi^{\dagger}\psi\psi, \eqno(1)$$ where
$H$ is the hamiltonian,
$N$ is the number operator, and $\mu$ is the chemical potential [8,9].
Here $T = -\hbar^2\nabla^2/2m$ is the
kinetic energy, $U({\bf r})$ is the external confining potential, and the
short-range interatomic two-body
potential has been approximated by a pseudopotential with an $s$-wave
scattering length  $a$.  The presence
of Bose condensation implies that the field operator has a macroscopic
ensemble average
$\langle \psi({\bf r})\rangle\equiv \Psi({\bf r})$,  identified as the (in
general,
temperature-dependent) condensate wave function. For a dilute system at low
temperature, most of the
particles are in the condensate, and the {\it deviation}\/ operator
$\phi({\bf r}) \equiv \psi({\bf r}) -
\Psi({\bf r})$ is treated as small (by definition, $\langle \phi({\bf
r})\rangle = 0$).  An expansion of  $K$
through second order in these small field amplitudes immediately yields $K
\approx K_0 + K'$, where
$$\eqalignno{K_0 &= \int dV\,\Psi^*(T + U - \mu)\Psi + 2\pi a\hbar^2m^{-1}\int
dV\,|\Psi|^4,&(2a)\cr
K' &= \int dV\,\phi^{\dagger}(T + U -\mu)\phi + 2\pi a\hbar^2m^{-1}\int
dV\,(4|\Psi|^2\phi^{\dagger}\phi
+ \Psi^2\phi^{\dagger}\phi^{\dagger} + \Psi^{*2}\phi\phi),&(2b)\cr}$$
and the first-order contribution vanishes because the first variation of
$K_0$ provides the nonlinear
Hartree  equation for the condensate wave function [10,11]
$$(T + U - \mu)\Psi + 4\pi a\hbar^2m^{-1} |\Psi|^2\Psi = 0.\eqno(3)$$
In addition, the ensemble average of the total number operator $N \equiv
N_0 + N'$
 determines the temperature-dependent number of particles in the condensate
$N_0 = \int dV\,|\Psi|^2$  and
in the excited states
$N' = \int dV\,\langle\phi^{\dagger}\phi\rangle$.  The
Bogoliubov approximation assumes that $N' <<N_0$,  thus neglecting  terms
of third and fourth order in the
deviation operators; this assumption clearly fails sufficiently close to
the onset temperature $T_c$,
since
$N_0(T_c)$ vanishes.

The first step is to determine the condensate wave function $\Psi$, which
then provides a static
interaction potential for the low-lying excitations.  Although the actual
experimental traps are anisotropic
[1,2], it is simplest to consider an isotropic three-dimensional harmonic
potential
$U({\bf r}) = {1\over 2}m\omega^2 r^2$,  with a characteristic oscillator
length $d_0 = \sqrt{\hbar/m\omega}$
(the effect of the anisotropy can be treated in perturbation theory).
Following Baym and Pethick [7],  I use a Gaussian trial function
$\Psi(r) = C\,\exp(-{1\over 2} r^2/d^2)$, where $C$ is a real normalization
constant and the length scale $d$
serves as the variational parameter.  Substitution into Eq.~(2$a$) yields
the variational quantity
$$K_0(\mu, d) = {\textstyle{1\over 2}}\pi^{3/2}\hbar\omega
\big[C^2({\textstyle{3\over 2} d\,d_0^2 + {3\over
2}}d^5d_0{}^{-2}) + C^4\sqrt{2}\pi a d^3d_0{}^2\big] -\mu
C^2\pi^{3/2}d^3.\eqno(4)$$
The thermodynamic identity [8] $\partial K_0/\partial \mu = -N_0$ fixes the
normalization $C^2 =
N_0/\pi^{3/2}d^3$, and the corresponding  energy becomes
$$E(\lambda)\equiv K _0 + \mu N_0 = {\textstyle{1\over 2} N_0\hbar\omega
\big[{3\over
2}}(\lambda^2 +\lambda^{-2})  + \sigma \lambda^3\big],\eqno(5)$$
where $\lambda \equiv d_0/d$ sets the spatial dimension of the spherical
condensate, and
$$\sigma \equiv \sqrt{2/\pi}\,( N_0a/d_0)\eqno(6)$$  is a dimensionless
parameter that characterizes the
relative strength of the interparticle energy (note that $\sigma$ is
proportional   to the parameter
$\zeta^5 \equiv 8\pi N_0a/d_0 = \sqrt{32\pi^3}\,\sigma$ defined in
Ref.~[7]).  In Eq.~(5), the
three terms  represent the kinetic energy, the confining energy, and the
interparticle
energy, respectively.

It is clear by inspection that $E(\lambda)$ becomes large for $\lambda\to
0$ (large $d/d_0$) because of the
spatial confinement in the harmonic potential, but the detailed behavior
for large $\lambda$ (small
$d/d_0$) depends on the value of the parameter $\sigma$.  In the absence of
the interparticle interaction
($\sigma = 0$), the minimum of Eq.~(5) occurs at $\lambda = 1$.  For any
positive $\sigma$ (repulsive
scattering length with $a > 0$), the cubic term eventually dominates for
$\lambda \to \infty$, and the local
minimum of
$E(\lambda)$ remains absolutely stable for all $\sigma \ge 0$.  For
negative $\sigma$ (attractive
scattering length with $a < 0$), however, the function
$E(\lambda)$  diverges to $-\infty$ for $\lambda \to\infty$,  and the local
minimum
disappears entirely at some critical negative value  $-|\sigma_c|$,
signaling the onset of an
instability.

        The condition $E'(\lambda_0)= 0$ determines the location of the
minimum, which satisfies the polynomial
equation $1 = \lambda_0^4 + \sigma\lambda_0^5$.    For $|\sigma|<<1$, the
root is given approximately by
$\lambda_0 \approx 1 - {1\over 4}\sigma$, with the corresponding  energy
$E_0 \approx {3\over 2}
N_0\hbar\omega(1 + {1\over 3}\sigma)$.  As expected, a small repulsive
(attractive) scattering length
 expands (contracts) the overall condensate size and raises (lowers) the
overall  energy.  For large
positive $\sigma$, it is straightforward to show that $\lambda_0 \approx
\sigma^{-1/5} -{1\over
5}\,\sigma^{-1}$, with
$E_0 \approx {5\over 4}N_0\hbar\omega\sigma^{2/5}$, as found in Ref.~[7].

        The situation is very different for negative $\sigma$,  because
$E(\lambda)$ no longer has a global
minimum, and even the local minimum  disappears at the critical values
$\lambda_c$ and $\sigma_c$ determined
by the simultaneous conditions
$E'(\lambda_c)~=~E''(\lambda_c)~=~0$.  An elementary calculation yields the
values
$$\lambda_c = 5^{1/4}\approx 1.495,\qquad\hbox{and}\qquad \sigma_c =
-{{4\over 5\lambda_c} =
-{4\over 5^{5/4}}\approx -0.535},\eqno(7)$$
so that the  interactions reduce the critical condensate size parameter
$d_c~\approx~0.669d_0$
relative to that of the bare trap;  the corresponding variational  energy
at the onset of
the instability is
$E_c\approx {1\over 2}\sqrt{5}\,N_0\hbar\omega$.  This calculation suggests
that the  energy becomes
unbounded from below for
$\sigma < \sigma_c$ through the disappearance of  the local stable minimum
rather than through the
onset of  negative  energy per particle.  Since this variational
calculation is merely an upper bound on the
 energy, the actual instability may well occur for less negative scattering
lengths, and, indeed,
numerical analysis from Ref.~[5] gives the value $\sigma_c \approx - 0.457$
for the vanishing of the
ground-state energy per particle;  the $\approx 15\%$ difference in these
values can be taken to
characterize the accuracy of  the variational estimate.

 As noted in Ref.~[2], this estimate predicts a
maximum condensate number
$N_0~\approx~1440$ for the  parameters ($a\approx -1.44$ nm and $d_0
\approx 3.13\;\mu\rm m$) appropriate
to the $^7$Li experiment.  This value is an order of magnitude less than
the total number  of trapped $^7$Li
atoms reported in Ref.~[2].  Even at zero temperature, however, the
nonzero interparticle potential ensures that $N_0(T = 0)< N$, and
finite-temperature excitation of
quasiparticles (see below)  further  reduces  $N_0(T)$ below $N$; hence it
is unclear whether this
discrepancy represents a failure of the Bogoliubov description. Although
the effect of three-body clusters
has recently been investigated [12], the small value of $|a|$ relative to
the interparticle spacing suggests
that two-body contributions dominate the physics of this many-body problem.

The next step is to  consider the noncondensate, which is  described by
the boson fields $\phi$ and $\phi^{\dagger}$. Since $K'$ in Eq.~(2$b$) is a
quadratic form in these field
operators, it can be diagonalized by a canonical transformation [9], as in
the original work of
Bogoliubov [3] for a uniform condensate.  Assume that the condensate wave
function has
the general form
$\Psi({\bf r}) = \sqrt{N_0}\,e^{iS({\bf r})}\,f({\bf r})$,
where the real amplitude function $f$ is normalized ($\int dV \,|f|^2 =
1$), and the phase $S$
produces a particle current ${\bf j} = \hbar N_0|f|^2m^{-1}\vec\nabla S$,
as found, for example, in a singly
quantized vortex [11,13,14],  where the appropriate $S$ is  the  angle in
cylindrical polar coordinates.
Define the linear  transformation [9]
$$\eqalign{\phi({\bf r}) &= e^{iS({\bf
r})}\,\mathop{{\sum}'}_j\big[u_j({\bf r})\,\alpha_j - v_j^*({\bf
r})\,\alpha_j^{\dagger}\big],\cr
 \phi^{\dagger}({\bf r}) &= e^{-iS({\bf
r})}\,\mathop{{\sum}'}_j\big[u_j^*({\bf r})\,\alpha_j^{\dagger} -
v_j({\bf r})\,\alpha_j\big],\cr}\eqno(8)$$
where the primed sum means to omit the condensate mode from the sum.  Here,
the ``quasiparticle''
operators
$\alpha_j$ and $\alpha_j^{\dagger}$ obey the usual boson commutation relations
$[\alpha_j,\alpha_k^{\dagger}]~=~\delta_{jk}$, $ [\alpha_j,\alpha_k] =
[\alpha_j^{\dagger},\alpha_k^{\dagger}] = 0$, ensuring that the
transformation is canonical, and the
wave functions $u_j$ and $v_j$ are chosen to satisfy the coupled ``Bogoliubov''
equations
$$\eqalign{Lu_j -4\pi a\hbar^2m^{-1}|\Psi|^2v_j &= E_ju_j,\cr
L^*v_j -4\pi a\hbar^2m^{-1}|\Psi|^2u_j &= -E_jv_j,\cr}\eqno(9)$$
where $L = -(2m)^{-1}\hbar^2(\vec \nabla + i\vec\nabla S)^2 +U - \mu + 8\pi
a\hbar^2m^{-1}|\Psi|^2$ is a
hermitian operator.  It is easy to verify that the eigenvalues $E_j$ are
real and that the
eigenfunctions obey the normalization $\int dV\,(u_j^*u_k - v_j^*v_k) =
\delta_{jk}$. Substitution of
Eq.~(8) into Eq.~(2b) yields the simple and physical result [9]
$$K' = -\mathop{{\sum}'}_jE_j\int dV \,|v_j|^2 +
\mathop{{\sum}'}_jE_j\,\alpha_j^{\dagger}\alpha_j,\eqno(10)$$
so that the canonical transformation indeed
diagonalizes the operator $K'$.  In addition, if $u_j$ and $v_j$ are a
solution with energy $E_j$, then
the pair $v_j^*$ and $u_j^*$ are also a solution with energy $-E_j$;  since
the quasiparticle number operator
$\alpha_j^{\dagger}\alpha_j$ has nonnegative integral eigenvalues, it is
necessary to take $E_j \ge 0$.
Finally, Eq.~(9) also has the solution $u_0 = v_0 = f$ with $E_0 = 0$,
verifying that the Bose condensation
does  occur in the lowest self-consistent single-particle mode.  Although
these equations are easily
rewritten in terms of two-component vectors (see, for example, pp.~477 and
501 of Ref.~[8]), such formalism is
unnecessary here.

The structure of $K'$ in Eq.~(10) leads to a very simple description of the
equilibrium  states of the  condensed Bose system. The quasiparticle ground
state $|{\bf 0}\rangle$
satisfies the condition
$\alpha_j|{\bf 0}\rangle = 0$ for all $j\neq 0$, and the excited states
follow by applying arbitrary number
of quasiparticle creation operators $\alpha_j^{\dagger}$ to $|{\bf
0}\rangle$. In addition, the well-known
properties of these harmonic-oscillator operators mean that the
low-temperature properties are determined
entirely by the eigenvalues and eigenfunctions of the Bogoliubov equations
(9).  If $\langle\cdots\rangle
\equiv {\rm Tr} [\,\cdots \,\exp(-\beta K')\,]/{\rm Tr}[\,\exp(-\beta
K')\,]$  denotes a self-consistent
ensemble average at temperature $T~=~(k_B\beta)^{-1}$, then the only
nonzero averages of one-  or
two-quasiparticle operators are
$\langle\alpha_j^{\dagger}\alpha_k\rangle =
\langle\alpha_k\alpha_j^{\dagger}\rangle - \delta_{jk} =
\delta_{jk}f_j$,
where $f_j \equiv [\exp(\beta E_j)-1]^{-1}$ is the usual Bose-Einstein
distribution function. For example
[9], the total number density $n({\bf r})$ has a condensate contribution
$n_0({\bf r}) = |\Psi({\bf r})|^2$
and a noncondensate contribution
$$n'({\bf r}) = \mathop{{\sum}'}_j\big[f_j\,|u_j({\bf r})|^2  + (1 +
f_j)\,|v_j({\bf r})|^2\big],\eqno(11)$$
where the condition $N = N_0(T) + \int dV\,n'({\bf r}) $ determines the
temperature-dependent condensate
fraction $N_0(T)/N$.

 In the present case of a spherical
condensate in a spherical confining potential $U(r)$, where $\Psi(r) =
\sqrt{N_0}\,f(r) $ satisfies Eq.~(3),
the  Bogoliubov equations simplify greatly because the states can be
characterized by the usual
angular-momentum quantum numbers ($l,m$) associated with the spherical
harmonics $Y_{lm}$, along with a radial
quantum number $n$.  Given a solution for $\Psi(r)$, standard
numerical techniques can determine the eigenvalues
$E_{nlm}$ and  associated radial eigenfunctions $u_{nlm}(r)$ and
$v_{nlm}(r)$ [6]. In order to gain more
physical insight, however, it is valuable to consider a special limiting
case in which the kinetic energy of
the condensate wave function is negligible compared to the confining energy
and the repulsive interparticle
interaction energy.  As discussed in [7] (see also Refs.~[4] and [15])
this condition holds for a harmonic
confining potential when the dimensionless parameter
$\sigma = \sqrt{2/\pi}\,(N_0a/d_0)$ from Eq.~(6) is large and positive,
because the kinetic energy is then of
order $\sigma^{-4/5}$ relative to the other two contributions.  As a
result, the Hartree equation (3) for
the condensate wave function then has the simple solution
$$4\pi a\hbar^2m^{-1}|\Psi(r)|^2 = \big[\,\mu -
U(r)\,\big]\,\theta\big[\,\mu - U(r)\,\big],\eqno(12)$$
where $\theta (x)$ denotes the unit positive step function. If  the
oscillator length $d_0$ is used to
scale dimensionless lengths, the normalization condition on
$\Psi$ then yields the   radius $R$ of the spherical condensate
$${R}^{5} = 15{N_0a/ d_0} = 15 (\pi/ 2)^{1/2}\,\sigma,\eqno(13)$$
with chemical potential  given by $\mu = {1\over 2}\hbar\omega R^2$.
Although this approximation
clearly fails in the immediate vicinity of  the condensate surface (see,
for example, Fig.~1 of Ref.~[4]),
 its use in the Bogoliubov equations leads to only a small error in the
limit $\sigma >>1$.

A combination of Eqs.~(9) and (12) yields the following coupled eigenvalue
equations
$$\eqalign{\big[D_x + V(x)\big]\,u_{nl}(x) - V_<(x)\,v_{nl}(x) &=
\epsilon_{nl}u_{nl},\cr
- V_<(x)u_{nl}(x) +\big[D_x +V(x)\big]\,v_{nl}(x)  &=
-\epsilon_{nl}v_{nl},\cr}\eqno(14)$$
 where $x=r/R$ and $\epsilon_{nl} = 2E_{nl}/\hbar\omega$. Here,
$$D_x \equiv -{1\over R^2}\bigg[{1\over x^2}{d\over dx}x^2{d\over dx}
+{l(l+1)\over x^2}\bigg] \eqno(15a)$$
is the kinetic-energy operator, and $V(x) \equiv V_<(x) + V_>(x)$ is the
potential energy, where
$$V_<(x) = R^2\,(1-x^2)\,\theta(1-x^2),\quad\hbox{and}\quad V_>(x) =
R^2\,(x^2-1)\,\theta(x^2-1)\eqno(15b)$$ are both positive.
  Apart from the coupling between $u$ and
$v$, which occurs only for
$x<1$ through  $V_<(x) $, these equations  look like those for  radial
eigenstates with orbital angular
momentum
$l$ in an isotropic repulsive  potential
$V(x) = R^2\,|1 - x^2|$, which has a central peak  at the origin, vanishes
linearly at $x = 1$, and rises
quadratically for $x >> 1$.  Thus the low-lying eigenfunctions are expected
to be ``surface'' modes localized
in the vicinity of the condensate surface at $x = 1$.

 In principle, these coupled differential equations can be solved
numerically, but   more physical
insight comes from recognizing that they have a variational basis.  If
${\cal U}(x)$ denotes a  two
component vector with elements $u(x)$ and $v(x)$, then Eq.~(14) has a
matrix representation
$$\big[{\cal D}_x +{\cal V}(x)\big]{\cal U}(x) = \epsilon \tau_3{\cal
U}(x),\eqno(16)$$
where ${\cal D}_x = D_x \,{\bf 1} $, ${\cal V}(x) = V(x)\,{\bf 1}+ V_<(x)
\,\tau_1$,  $\bf 1$ is the $2\times
2$ unit matrix, and $\tau_i$ are the familiar $2\times 2$ Pauli matrices.
It follows immediately that the
 variational quantity
$$\epsilon_{0l} \le {\int_0^{\infty}x^2\,dx \,{\cal U}^{\dagger}(x)
\,\big[{\cal D}_x + {\cal
V}(x)\big]\,{\cal U}(x)\over \int_0^{\infty}x^2\,dx \,{\cal U}^{\dagger}(x)
\,\tau_3\,{\cal U}(x)}\eqno(17)$$
 provides an upper bound on the lowest eigenvalue $\epsilon_{0l}$ for each
separate $l$.  As a very simple
model, take
$${\cal U}(x) = \pmatrix{\cosh \chi\cr\sinh \chi\cr}\,g(x),\eqno(18)$$
with $\int_0^{\infty}x^2\,dx\,|g(x)|^2 = 1$.  Substitution into Eq.~(17)  gives
$$\epsilon_{0l} \le A \cosh 2\chi - B\sinh 2\chi,\eqno(19)$$
 where
$$A = \int_0^{\infty}x^2\,dx\, g(x)^*\,\big[\,D_x + V(x)\,\big]\,g(x) \quad
\hbox{and}\quad B =
\int_0^1x^2\,dx\, g(x)^*\,V_<(x)\, g(x).\eqno(20)$$  Minimization with
respect to $\chi$ yields the
condition $\tanh 2\chi = B/A$, with
$$\epsilon_{0l}\le \sqrt{A^2 - B^2}.\eqno(21)$$

If $g$ also depends on some parameters, they can be varied to find the
minimum of Eq.~(21).  For example,
take $g(x) \propto x^{\gamma}\,\theta(1 - x) + x^{-\gamma-1}\,\theta(x -
1)$.  The integrals in
$A$ and $B$ are easily evaluated (with an integration by parts in the case
of $A$), as is the normalization
integral, and the minimum with respect to $\gamma$ was found numerically
for several different values of $l$
and
$R$.  Since the description is expected to hold best for larger $R$, only
the case of $R = 5$ will be
considered in detail, corresponding to a value $\sigma \approx 168$ for the
dimensionless parameter
$\sigma$ that determines the relative importance of the kinetic
contribution in the energy balance for the
condensate wave function.  To a good approximation, the 11 lowest eigenvalues
$\epsilon_{0l}$ for
$l = 0,
\cdots, 10$ can be fit to a quadratic polynomial
$\epsilon_{0l}
\approx 5.83 + l/25.05 +l^2/25.25$, which effectively has the intuitive
form $\epsilon_{0l} \approx
\epsilon_{00} + l(l+1)/R^2$ of a radial zero-point energy $\epsilon_{00}$
plus  rotational energy
$l(l+1)/R^2$ of a rigid rotor; similar calculations for other values of $R$
show that $\epsilon_{00}(R)$
depends only weakly on
$R$, as expected from the form of the potential $V(x)$. For comparison, the
eigenvalues of the bare confining
harmonic potential (here assumed isotropic) are
$4n + 2l + 3$ in the same units of
${1\over 2}\hbar\omega$ [16]. The most striking conclusion here is that the
low-lying elementary excitations
of the Bose condensate for relatively large condensate radius $R$ and
condensate number $N_0$ should have a
 rotational band of states (those for
$n = 0$ and
$l = 0, 1, 2,
\dots$) lying somewhat above the lowest state of the bare confining
potential.  With the values
$a
\approx 10$ nm and
$d_0\approx 1.4
{}~\mu\rm m$, which are appropriate for the experiment in Ref. [1], the
radius $R = 5$ corresponds to a
condensate number
$N_0
\approx 29,000$, about an order of magnitude larger than the value
estimated in Ref.~[1].  In Ref.~[6], where
the Bogoliubov equations were solved numerically  in the presence of the
``exact'' condensate density, the
corresponding dimensionless parameter is
$\sigma \approx 9.7$;  thus it is not surprising that their eigenvalues
(Ref.~[6] reports only those for $l
\le 4$) differ considerably from the simple rigid-rotor form given above.

The particle density operator $\rho({\bf r}) =\psi^{\dagger}({\bf
r})\psi({\bf r})$ plays a
central role in the response of a physical system to external
perturbations.  In the present case of
 a dilute Bose condensate, the noncondensate density $\rho' \equiv \rho -
|\Psi|^2$   has an unusual and
characteristic form
$$\rho' \approx  \Psi^*\phi + \Psi\phi^{\dagger}\eqno(22)$$ that follows
from the
Bogoliubov approximation introduced below Eq.~(1).    Since the operators
$\phi({\bf r},t)$ and
$\phi^{\dagger}({\bf r},t)$ oscillate harmonically at the frequencies given
by the eigenvalues of the
Bogoliubov equations, Eq.~(22) shows that the normal modes of the
condensate can be identified as density
(compressional) waves.  In particular,  the noncondensate part of the
density-density correlation function
becomes simply a correlation function of the deviation operators, given by
$$\eqalign{{\langle \rho'({\bf r},t)\rho'({\bf r'},0)\rangle\over|\Psi({\bf
r})\Psi({\bf r'})|}  \approx
\mathop{{\sum}'}_j\big\{(1 &+ f_j)\,\big[u_j({\bf r})-v_j({\bf
r})\big]\,\big[u_j^*({\bf r'})-v_j^*({\bf
r'})\big]\,e^{-iE_jt/\hbar} \cr
&+f_j\,\big[u_j^*({\bf r})-v_j^*({\bf r})\big]\,\big[u_j({\bf r'})-v_j({\bf
r'})\big]\,e^{iE_jt/\hbar}\big\};
\cr}\eqno(23)$$
thus a measurement of the frequency spectrum of density oscillations (for
example, by studying the
resonant response to small modulations of the trapping potential) would
directly characterize the eigenvalues
$E_j$.

I am grateful to M. Kasevich and M. Levenson for  stimulating and
invaluable discussions.   This work
 was
supported in part by the National Science Foundation, under Grant No.~DMR
94-21888.

\vskip 2cm
\centerline{References}
\noindent [1] M.~H.~Anderson, J.~R.~Ensher, M.~R.~Matthews, C.~E.~Wieman,
and E.~A.~Cornell,
Science {\bf 269},198 (1995).

\noindent [2] C.~C.~Bradley, C.~A.~Sackett, J.~J.~Tollett, and R.~G.~Hulet,
Phys.~Rev.~Lett.~{\bf 75}, 1687
(1995).

\noindent [3] N. N. Bogoliubov, J.~Phys.~(USSR) {\bf 11}, 23 (1947).

\noindent [4] M. Edwards and K. Burnett, Phys.~Rev.~A {\bf 51}, 1382 (1995).

\noindent [5] P.~A.~Ruprecht, M.~J.~Holland, K.~Burnett, and M.~Edwards,
Phys.~Rev.~A{\bf 51}, 4704 (1995).

\noindent [6] M.~Edwards, P.~A.~Ruprecht, K.~Burnett, and C.~W.~Clark,
(unpublished).

\noindent [7]  G.~Baym and C.~Pethick (unpublished).

\noindent [8] See, for example, A.~L.~Fetter and J.~D.~Walecka, {\it
Quantum Theory of Many-Particle Systems}
(McGraw-Hill, NY, 1971), Chaps.~2, 10, and 14.

\noindent [9] A.~L.~Fetter, Ann.~Phys.~(NY) {\bf 70}, 67 (1972).

\noindent [10] V.~L.~Ginzburg and L.~P.~Pitaevskii,
Zh.~Eksp.~Teor.~Fiz.~{\bf 34}, 1240 (1958);
Sov.~Phys.--JETP {\bf 7}, 858 (1958).

\noindent [11]  E.~P.~Gross, Nuovo Cimento {\bf 20}, 451 (1961);
J.~Math.~Phys.~{\bf 46}, 137 (1963).

\noindent [12]  B.~D.~Esry, C.~H.~Greene, Y.~Zhou, and C.~D.~Lin (unpublished).

\noindent [13]  L.~P.~Pitaevskii, Zh.~Eksp.~Teor.~Fiz.~{\bf 40}, 646 (1961);
Sov.~Phys.--JETP {\bf 13}, 451 (1961).

\noindent [14]  A.~L.~Fetter, Phys.~Rev.~{\bf 138} A429 (1965); {\bf 138}
A709 (1965); {\bf 140} A452
(1965).

\noindent [15]  D.~A.~Huse and E.~D.~Siggia, J.~Low Temp.~Phys.~{\bf 46},
137 (1982).

\noindent [16]  See, for example, Ref.~[8], pp.~508-511.
\bye